\documentclass[preprint,aps]{revtex4}

\usepackage{graphicx}
\usepackage{dcolumn}
\usepackage{bm}

\newcommand{\be}{\begin{equation}}  
\newcommand{\ee}{\end{equation}}  
\newcommand{\bear}{\begin{eqnarray}}  
\newcommand{\eear}{\end{eqnarray}}  
\newcommand{\ba}{\begin{array}}  
\newcommand{\ea}{\end{array}}

\newskip\humongous \humongous=0pt plus 1000pt minus 1000pt

\newif\ifdtup

\def\oldreffmt#1{\rlap{[#1]} \hbox to 2\parindent{}}

\def\figfmt#1{\rlap{Figure {#1}} \hbox to 1in{}}  
  
\def\ie{\hbox{\it i.e.}{}}	\def\etc{\hbox{\it etc.}{}}  
\def\eg{\hbox{\it e.g.}{}}

\def\Tr{\mathop{\rm Tr}}

\def\slash#1{#1\!\!\!/\!\,\,}  
\def\beq{\begin{equation}}  
\def\eeq{\end{equation}}  
\def\bea{\begin{eqnarray}}  
\def\eea{\end{eqnarray}}  
\def\half{\frac{1}{2}}  
  
\def\bq{\begin{quote}}  
\def\eq{\end{quote}}

\def\half{\frac{1}{2}}       

\relax  

\newdimen\tdim  
\tdim=\unitlength  
\def\bar{\overline}

\begin{document}

\preprint{FERMILAB-PUB-06-046-T}

\vskip 0.4in
\title{Exact Equivalence of the \\
$D=4$ Gauged Wess-Zumino-Witten Term \\ 
and the $D=5$ Yang-Mills Chern-Simons Term\\
}

\author{Christopher T. Hill }

\affiliation{
 {{Fermi National Accelerator Laboratory}}\\
{{\it P.O. Box 500, Batavia, Illinois 60510, USA}}
}%

\date{\today}

\begin{abstract} 

We derive the full  Wess-Zumino-Witten term
of a gauged 
chiral lagrangian in $D=4$ by starting from a pure Yang-Mills theory
of gauged quark flavor in a flat, compactified $D=5$. 
The theory is compactified such
that there exists a $B_5$ zero mode, and supplemented with
quarks that are ``chirally delocalized'' with $q_L$ ($q_R$) on the left (right)
boundary (brane). The theory then necessarily contains a Chern-Simons term
(anomaly flux) to cancel the fermionic anomalies on the boundaries. 
The constituent quark mass represents chiral symmetry breaking
and is a bilocal operator in $D=5$ of the form: $\bar{q}_LWq_R+h.c$, 
where $W$ is the Wilson line 
spanning the bulk, $0\leq x^5 \leq R$, 
and is interpreted as a chiral meson field,
$W=\exp(2i\tilde{\pi}/f_\pi)$, where $f_\pi \sim 1/R$.  
The quarks are integrated out, yielding
a Dirac determinant which takes the form of
a ``boundary term'' (anomaly flux return), and is equivalent to
Bardeen's counterterm that connects consistent and covariant anomalies. 
The  Wess-Zumino-Witten term then emerges straightforwardly,
from the Yang-Mills Chern-Simons term, plus boundary term.
The method is systematic and allows generalization of the 
Wess-Zumino-Witten term to theories
of extra dimensions, and to express it in alternative and
more compact forms. We give a novel
form appropriate to the case of (unintegrated) massless
fermions.
\end{abstract}

\email{hill@fnal.gov}

\maketitle

\section{\bf Introduction}

In this paper we derive the full  Wess-Zumino-Witten term \cite{wess,witten}
of a gauged 
chiral lagrangian in $D=4$ by starting from an $SU(N_f)$ Yang-Mills theory
of gauged quark flavor in compactified $D=5$. 
The Yang-Mills gauge fields
propagate in the bulk,
with chiral quarks attached to boundaries (branes) 
located at $x^5=0$ and $x^5=R$.
The quarks are chirally delocalized, \ie, their $SU(N_f)$ flavor
anomalies are nonzero on their respective boundaries, but would
otherwise cancel if the boundaries
were merged.  

The  boundary conditions
on the Yang-Mills gauge fields, $B_A$, 
are subject to a minimal set of constraints:
{\em (1)} there exists a massless physical $B_5$ zero mode
that can be identified with mesons and, {\em (2)} 
there exists a tower of KK-modes
of the spin-1, $B_\mu^n$ that is sufficiently rich such that 
independently valued combinations of these fields, 
exist on the boundary branes, 
$B_L = B_\mu(x^\mu, 0)$ and $B_R= B_\mu(x^\mu, R)$. 
With judicious choices of compactification schemes, such as $S_1$ or
``flipped orbifolds,'' one can imitate the spectrum of QCD, but 
this need not be specified presently. 

Much of what we 
say will apply to any theory of new physics in extra dimensions that
satisfies {\em (1)} and {\em (2)} with chiral delocalization.  
The reason is that the results
are largely homological, \ie, they are determined
at the boundary of the bulk,  as the integrals over
the bulk involving the lower KK-modes are mostly exact expressions.
In addition, some inexact (bulk integral) components are generated,
reflecting new interactions amongst
KK-modes that are contained in the Chern-Simons term \cite{hill}.

With the chiral quarks attached to the boundaries, 
$\psi_L$ at $x^5=0$ and $\psi_R$ at $x^5=R$ respectively,
a ``constituent quark mass term'' 
is introduced of the form $m\bar{\psi}_L W\psi_R + h.c.$.
Here $W$ is the Wilson line that spans the gap between the
boundary branes, and represents the dynamical chiral condensate
of the theory.  The Wilson line is identified with
the chiral field of mesons:
\beq
W(x^\mu) = P\exp\left( -i\int_0^{R} dx^5 B_5(x^\mu,x^5)\right) 
\equiv \exp(2i\tilde{\pi}(x^\mu)/f_\pi)
\eeq
where $\tilde{\pi} = \pi^a\lambda^a/2$ and $f_\pi = 93$ MeV.  
This is the reason for
having a $B_5$ zero mode, since we desire
that the $\tilde{\pi}$ is
physical, and not eaten by a KK-mode. 
In any imitation of QCD 
chiral dynamics by an extra dimension,
chiral symmetry breaking is intrinsically related to 
the compactification scale,
\ie, $f_\pi \sim 1/R$.

The chiral delocalization of the quarks implies
the failure of anomaly cancellation on
each brane.  This mandates a
Chern-Simons term spanning the bulk and terminating
on the two quark branes.
The fermionic anomalies \cite{jackiw,adler,bardeen}
on the boundary branes under $SU(N_f)$ flavor transformations must
be cancelled by the anomalies that arise on the boundaries from
the Chern-Simons term under the same gauge transformation.
This cancellation condition determines the coefficient of
the Chern-Simons term. 
We must use the {\em consistent anomalies} of the
fermions,
\ie, the anomalies that come directly from Feynman diagrams
(see \cite{hill}). 
We see below that 
the Chern-Simons anomaly has precisely the same
form as the consistent anomaly of the fermion current
\cite{bardeen}.  We integrate out
the quarks in taking the limit of large $m$.  This
generates, through the Dirac determinant,
a {\em boundary term}, which is the effective interaction
amongst the gauge fields on the boundaries, $B_L$ and $B_R$.
It arises from triangle and box loops, and has a structure 
identical to that of the  counterterm that maps consistent anomalies
into covariant anomalies, as introduced by Bardeen 
(\cite{bardeen}, eq.(45)).

Note that, as a metaphor for the structure of
the theory, we can view
the Chern-Simons term as an ``anomaly flux''
that runs from one boundary to the other.  
We can likewise view
the boundary terms as an ``anomaly return flux.''
When added together, 
\beq
\label{tilS}
\tilde{S} = S_{CS} + S_{boundary}
\eeq
we have a total effective action, $\tilde{S}$, that 
contains no net anomalies and
generates the topological physics of the bosons
in a fully gauge invariant way.
The low energy effective theory, 
truncated on the $B_5$ zero mode and fermions, 
becomes a chiral lagrangian with the
symmetry $SU(N_f)\times SU(N_f)_R$. The $B_\mu$ modes can play the role
of fundamental gauge fields, or as
vector
and axial vector mesons, coupled to the mesons.  

We find in Section III that $\tilde{S}$ resolves 
into two classes of terms, 
homological surface terms, and bulk terms.
With this construction, we can straightforwardly
derive the full Wess-Zumino-Witten (WZW) term.  The 
leading  term, $S_{CS0}\sim \int\Tr(\pi d\pi d\pi d\pi d\pi)$, arises
immediately from this construction \cite{hill}, however it
is the leading term in an expansion
in mesons $\pi$.   The fully gauged WZW term, $S_{WZW}$, emerges
as the remaining set of exact boundary terms in $\tilde{S}$.  
In addition we have
an interaction term in the bulk, $S_{bulk}$, which generates
new interactions amongst the KK-modes from the
Chern-Simons term.  These new bulk interactions
were previously studied in detail for QED in \cite{hill},
and the basic procedure developed there is followed here.  
Our anomaly free action involves
cancellation of anomalies between $S_{bulk}$ and $S_{WZW}$.
A chiral lagrangian in $D=4$, such as QCD, has no bulk term, and
we would simply omit $S_{bulk}$, leaving an anomalous $S_{CS0}+S_{WZW}$
as the fully gauged Wess-Zumino-Witten term of the theory. 
Our results confirm the original analysis of Witten, \cite{witten}, 
in the finalized form given by Kaymakcalan, 
Rajeev and Schechter \cite{kay}, and
Manohar and Moore, \cite{man}.   We'll maximally conform to
the notation of \cite{kay} so that our final results are
directly comparable to their eq.(4.18).  

We emphasize that our procedure is significantly different from
the traditional way in which the full WZW term was originally
derived. In the standard approach, pioneered
by Witten, one  {\em starts with a chiral theory of  mesons} 
in $D=5$, and incorporates
a $D=5$ chirally invariant 
pionic interaction, $\Tr(d\pi d\pi d\pi d\pi d\pi)$, 
with a quantized coefficient. Upon descending to $D=4$, this yields
the $\Tr(\pi d\pi d\pi d\pi d\pi)$ as a boundary term. 
This is subsequently gauged, {\em a posteriori}, 
by performing gauge transformations,
observing new terms that are generated by the transformation,
and then compensating these with the addition of gauge
field and meson interactions. The main difference 
is that our present procedure 
begins with a pure Yang-Mills theory and
is {\em a priori} gauge invariant. The mesons are ``born''
as we descend by compactification
and identify a gauge field $B_5$ with $\tilde{\pi}$.
As a result the full WZW term has a larger gauge
symmetry involving the ``mesons'' together with the gauge fields.

We believe that the present analysis amplifies
the structure and significance of the full WZW term,
and it illustrates technically better ways to manipulate it, and 
implies that the WZW term is a more general gauge invariant object.
To illustrate this, we show in Section IV how to immediately write
down the WZW term in the case where the fermions have small masses,
and are not integrated out. This is a particularly instructive
example as to how the machinery of the full WZW term operates.

\section{The General Set-up}

\subsection{Fermionic and Gauge Kinetic Terms}

\begin{figure}[t]  
\vspace{4.5cm}  
\includegraphics{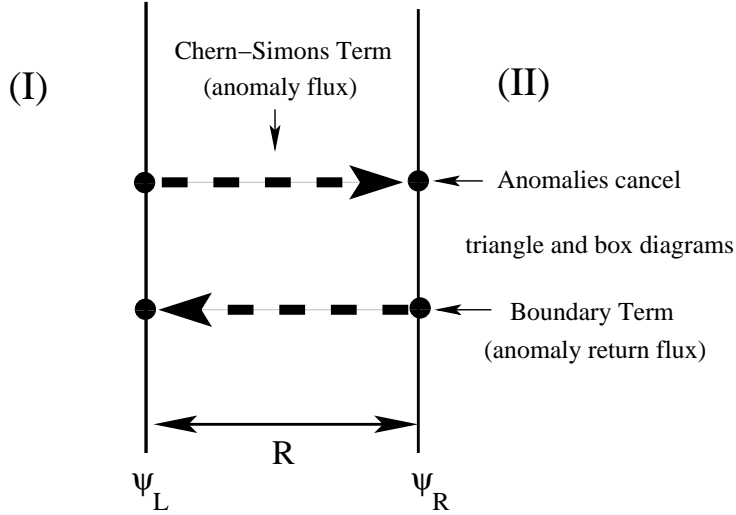}  
\vspace{3.5cm}  
\caption[]{
\addtolength{\baselineskip}{-.3\baselineskip}  
Orbifold with split, anomalous fermions (quarks).
$\psi_L$ ($\psi_R$) is attached to the  $D=4$ left-boundary
brane, $I$ (right-boundary brane,
$II$).  Gauge fields propagate in the $D=5$ bulk, which has a compactification
scale $R$. The fermions have a Wilson line mass term, $m\bar{\psi}_L
W\psi_R+h.c.$. The bulk contains a Chern-Simons term. The branes,
upon integrating out massive fermions in the large $m>>1/R$ limit,
yield a ``boundary term'' in the effective action which
takes the form of the negative of Bardeen's counterterm. The anomalies
from the Chern-Simons term cancel the anomalies from the triangle diagrams
on the respective branes so the overall theory is anomaly free. 
The Chern-Simons action plus boundary term yield the Wess-Zumino-Witten term
and a residual bulk interaction amongst KK-modes.}  
\label{dirac2}  
\end{figure}  

Consider a generic $D=5$, $SU(N)$ Yang-Mills  gauge theory compactified
on a physical interval $0 \leq x^5 \leq R$. 
We have vector potentials, $B_A^a(x)$ 
and coordinates $x^A$, where $(A=0,1,2,3,5)$,
the covariant derivative,
\beq
D_A = \partial_A - iB_A \qquad \qquad  B_A = B_A^aT^a ,
\eeq
where, \eg, $T^a=\lambda^a/2$ in the adjoint representation
of $SU(3)$.
The field strength is:
\beq
G_{AB} = i[D_A,D_B]
=\partial_A B_B-\partial_B B_A - i[B_A,B_B] .
\eeq
For completeness, the bulk Yang-Mills kinetic action is:
\bea
\label{kinetic}
S_0 & =& -\frac{1}{2\widetilde{g}^2}\int_0^R dy\int d^4 x \Tr G_{\mu\nu}G^{\mu\nu}
-\frac{1}{\widetilde{g}^2}\int_0^R dy\int d^4 x
\Tr G_{\mu 5}G^{\mu 5}.
\eea
where, by
examining the zero mode $B_\mu$, 
the physical coupling constant is $g^2 = \widetilde{g}^2 /R$.
Beyond this we don't have to be too specific.

We are interested presently in any choice of boundary conditions
such that: (1) the lowest
$B_5$  mode is massless and
physical, \ie, not eaten by any massive spin-1 modes; 
(2) we have a tower of KK-modes of $B_\mu$,
which may or may not have a massless zero mode, $B_\mu^0$.
The low lying KK-modes yield distinct fields
$B_L(x^\mu, 0)$ and $B_R(x^\mu, R)$ on the boundaries
at $y=0$ and $y=R$ respectively.  
There are various choices of boundary conditions for the gauge fields that
can lead to this, and can
even imitate QCD.  Compactification onto $S_1$, or
``flipped orbifolds'' \cite{hill}, can lead to the physics of interest.  
We will not discuss these cases presently, as only
the most general aspects of the KK-mode spectrum 
embodied in (1) and (2)  are required
for this analysis.

We introduce chiral quarks with $N=N_f$ flavors (and $N_c$ ungauged
colors) onto
the boundaries, $I$ and $II$, located respectively
at $x^5=0$ and $x^5=R$.
The fermionic matter action on the boundaries is:
\bea
S_{kinetic} & = &   \int_I d^4x \;\bar{\psi}_Li\slash{D}_L\psi_L 
+  \int_{II} d^4x \;\bar{\psi}_Ri\slash{D}_R\psi_R
\eea
where: 
\beq
D_{L\mu} = \partial_\mu - iB_\mu(x_\mu, 0)\; ,
\qquad \qquad D_{R\mu} = \partial_\mu - iB_\mu(x_\mu, R)
\qquad \qquad B_\mu = \frac{\lambda^a}{2}B^a_\mu
\eeq
The $B_\mu$ act upon the flavor indices,
and:
\beq
\psi_L = \frac{1-\gamma^5}{2} \psi \; , \qquad
\qquad
\psi_R = \frac{1+\gamma^5}{2} \psi \; .
\eeq
The $\psi_L$ and $\psi_R$
chiral projections are key ingredients of
the theory.
This structure can, of course, come
about if there is a thin domain wall (kink) at $x^5=0$ and an
anti-domain wall (anti-kink) at $x^5=R$, where $\psi_L$ and $\psi_R$
are then the fermionic zero modes. The $B_L = B_\mu(x_\mu, 0)$
($B_R = B_\mu(x_\mu,R)$) is the left (right) gauge field
and is just the bulk gauge field value on the brane at $x^5=0$ ($x^5=R$),
the sum over KK-modes on the respective
boundary, \eg, $B_R =\sum^n B^n_\mu(x^\mu, R)$.

We further introduce a ``constituent quark mass term''
of the form:
\beq
S_{mass} = \int\; d^4 x \; m \bar{\psi}_L W\psi_R + h.c.
\qquad \qquad W = P\exp\left(-i\int_0^{R} dx^5 B_5\right)
=\exp(2i\tilde{\pi}/f_\pi)
\eeq
We have defined the Wilson line
as the chiral meson field. 
At this stage the Wilson line contains all the modes of $B_5$,
and it is ambiguous to separate the zero-mode from non-zero modes.
This separation is gauge dependent, and we must define the ``meson''
fields, subsequent to some preparatory manipulations.

\subsection{Transforming to Axial Gauge, $B_5\rightarrow 0$}

Consider ``gauging away'' $B_5$, which also
zeros the Wilson line. 
We assume that there is no topological
invariant associated with the Wilson line, \ie, 
no nontrivial physical flux is
encircled by the line integral that would be an obstruction
to performing this gauge transformation. 
The transformation removes the Wilson line,
hence the ``mesons,'' from the quark mass term.

To implement this, we consider a Wilson line that runs
from brane $I$ into the bulk to a position $y$: 
\beq
\label{trans}
V(x^\mu, y) = P\exp\left(-i\int_0^{y} dx^5\; B_5^0(x^\mu, x^5) \right)
\eeq
and note the essential path-ordering. We now have:
\beq
i\partial_y V(x^\mu, y) =   V(x^\mu, y) B_5^0(x^\mu,y)
\qquad
i\partial_y V^\dagger(x^\mu, y) =  -B_5^0(x^\mu,y) V^\dagger(x^\mu, y) 
\eeq
Thus we can consider a gauge transformation:
\beq
\label{anom1}
\psi'_L = \psi_L ,\;\; \qquad 
\psi'_R = V(R) \psi_R \;\; 
\eeq
and:
\beq
\tilde{B}_A (x^\mu,y) = V ({B}_A +i  \partial_A) V^\dagger 
\eeq
hence:
\beq
\tilde{B}_\mu(x^\mu, y)  =
V ({B}_\mu + i\partial_\mu) V^\dagger
\qquad \qquad \tilde{B}_5(x^\mu, y) =  V ({B}_5 + i\partial_y) V^\dagger =  0
\eeq
The kinetic terms, mass term, and Wilson line thus transform as:
\beq
\bar{\psi}(i\slash{\partial} + \slash{B})\psi
= \bar{\psi}'(i\slash{\partial} + \slash{\tilde{B}})\psi'
\qquad  \bar{\psi}_LW \psi_R
= \bar{\psi}'{}_L\psi'{}_R
\qquad 
W\rightarrow  V(0) W V^\dagger(R) =  1.
\eeq
The field strengths transform covariantly, $G_{AB}\rightarrow
V G_{AB} V^\dagger$. Note that fermionic anomalies generated
by eq.(\ref{anom1}) will be cancelled by the Chern-Simons term.

The physical $\tilde{B}_\mu$ is now a tower
of spin-$1$ fields that have become comingled
with the mesons. Hence we wish to extract the 
$\tilde{\pi}$ meson fields.  
We now define a particular unitary matrix
of the form:
\beq
\label{uy}
\tilde{U}(y) \equiv \exp(2ih(y)\tilde{\pi}/Rf_\pi),\qquad\qquad \tilde{\pi} = \pi^a T^a
\eeq
where $y=0$ ($y=R$) is the left (right) boundary.  In this
expression $h(y)$ is {\em ab initio} any monotonic function that 
takes on values $h(0)=0$ and $h(R)=1$.  As we'll
see momentarily, $h(y)$ controls the longitudinal mixing 
of $\partial_\mu\tilde{\pi}$
with the pseudovector components of $B_\mu$ in $G_{\mu 5}$,
and  $h(y) = y/R$ is preferred if $\tilde{\pi} $ is
a pseudoscalar.
We note that $\tilde{U}$
requires no path ordering because $\tilde{\pi}$ is independent of
$y$. 

We then define the gauge field $A_\mu(x^\mu, y)$ by the redefinition:
\beq
\label{pull}
\tilde{B}_\mu(x^\mu, y) 
= \tilde{U}(y) (A_\mu(x^\mu, y) + i\partial_\mu ) \tilde{U}^\dagger(y)
\eeq
We emphasize that the definition of eq.(\ref{pull}) 
is {\em not a gauge transformation}
of the full $B_A$, but is only a {\em redefinition }
of the $\tilde{B}_\mu$. That is, we do not allow this
redefinition to act upon the fermions, 
so it does not reintroduce the Wilson line
into the mass term. The redefinition isolates the mesons from the gauge
fields in the first few terms of an
expansion in $\tilde{\pi}$,
and is sufficient to yield the WZW term below.

As an aside,
we note that 
the redefinition of eq.(\ref{pull}) does not compromise the $D=4$
gauge invariance
of the theory. If we perform a general gauge transformation
$\tilde{B}_\mu(x^\mu, y)\rightarrow V\tilde{B}_\mu(x^\mu, y)V^\dagger$
we will have an induced gauge transformation
$A_\mu \rightarrow Y (A_\mu + i\partial_\mu )Y^\dagger$ where
$\tilde{U}(y)^\dagger V \tilde{U}(y) =  Y$. The generators of $Y$ 
are $\tilde{U}(y)^\dagger T^a\tilde{U}(y)
= T_Y^a$ and satisfy the Lie algebra of $SU(N)$. This redefinition works
so long as we don't involve $\partial_y$ in the $Y$ transformation. 

The redefinition acts covariantly
as a gauge transformation on $G_{\mu\nu}$:
\beq
G_{\mu\nu}(\tilde{B}) = \tilde{U}(y)G_{\mu\nu}(A)\tilde{U}^\dagger(y)
\eeq
but not so on $G_{\mu 5}$, since we do not transform $\tilde{B}_5=0$.
We find:
\bea
\label{inc}
G_{\mu 5} & = & -\partial_y [\tilde{U}(y) (A_\mu + i\partial_\mu )
\tilde{U}^\dagger(y)]
\nonumber \\
& = & -({2ih'(y)}/{f_\pi})
\tilde{U}(y) (i\partial_\mu \tilde{\pi}+[\tilde{\pi}, A_\mu] )
\tilde{U}^\dagger (y)  - \tilde{U}(y) \partial_y A_\mu \tilde{U}^\dagger (y)
\eea
where $\partial_y (\tilde{U}(y)\partial_\mu 
\tilde{U}^\dagger(y))= (2ih'(y)/f_\pi)\tilde{U}(y)\partial_\mu\tilde{\pi} 
\tilde{U}^\dagger(y) $ is exact.

Thus, a $\Tr(G_{\mu 5})^2$ kinetic term contains:
\bea
\label{kin5}
-\frac{1}{\widetilde{g}^2}\int_0^R dy\int d^4 x
\Tr G_{\mu 5}G^{\mu 5} 
& = & \frac{4}{\widetilde{g}^2 f^2_\pi}\int_0^R dy\int d^4 x\Tr \left(
h'(y)\partial_\mu \tilde{\pi} -i h'(y)[\tilde{\pi}, A_\mu] - \frac{i}{2}f_\pi \partial_y A_\mu 
\right)^2 \nonumber \\
\eea
(note a minus sign from raising the $5$ index). Setting $A_\mu=0$
and performing the $y$ integral, (and we
assume $h(y)$ is normalized as $\int dy h'(y)^2 = z/R$),
we see the emergence of the
meson kinetic term, $\Tr(d\tilde{\pi})^2$ with
$f_\pi^2 = 4z/R\widetilde{g}^2 = 4z/R^2{g}^2$. 

$A_\mu = \sum_n A^n_\mu$, is a tower of KK-modes
containing all of the the non-zero-mode 
longitudinal components of the original
$B_5$, \ie, the  longitudinal spin degrees of freedom. These are the
Yang-Mills generalizations of ``Stueckelberg fields'', which
have the form, expanding in $B_5$,
 $A^n_\mu \sim B^n_\mu -\partial_\mu B^n_5/M^n + ...$.
Under gauge transformations, $\delta B_\mu^n\rightarrow = 
\partial_\mu \theta^n$
and $\delta B_5^n = \theta^n M^n$, the $A_\mu$ transform covariantly.
If the gauge transformation is a valid symmetry of the theory, then we
can always bring the massive modes into this covariant
form.

The wave-functions of the $A^n_\mu$ in 
$y$ are normalized by the kinetic term, $\int_0^R dy \Tr(G_{\mu\nu}^2)$.
For a typical compactification scheme on the interval
$[0,R]$ we may have an $A_\mu$ zero mode, 
with a flat wave-function, $A^0\sim \sqrt{\tilde{g}^2/R}$
and KK-mode excitations with
and $A^n\sim \sqrt{2\tilde{g}^2/R}\cos(n\pi y/R)$, where the $\tilde{g}$
factor rescales the $A^n$ to canonical normalization.
Setting $\tilde{\pi}=0$ in eq.(\ref{kin5})
we see that the mass terms for KK-modes are contained in the
usual term,
$\Tr \bigl(\partial_y A_\mu \bigr)^2$, and are computed in the mode
expansion. With the properly normalized $A_\mu^n$, we have $M^n = \sqrt{2}
\pi n/R$.

We also
have the longitudinal coupling of the mesons to the
vector potentials, contained in the term of the form $\Tr 
\bigl(h'(y)\partial^\mu \tilde{\pi} \partial_y A_\mu \bigr)$.
This requires a matching of $h(y)$ to the $y$ dependence
of the wave-function of
the KK-modes in  $A_{\mu}$ to establish the parity of
the $\pi$ in a consistent manner.  
Note that for the typical mode functions, the $1^-$, normal
parity $A_\mu^0$ zero mode (corresponding to the ``$\rho$-meson octet'')
has  $\partial_yA_\mu^0 =0$, thus it decouples from $\partial_\mu\tilde{\pi}$. 
On the other hand, the first KK-mode, $A^1_\mu $ corresponds to an abnormal
parity $1^+$ state (the ``$A^1$ octet''), and  $\partial_y A^1_\mu \sim 
\sin(\pi y/R)$.  
By requiring,
\beq
h'(y) =\frac{1}{R}
\qquad \qquad
h(y) = \frac{y}{R}.
\eeq 
we see that $\partial_\mu\tilde{\pi}$ will be orthogonal to all $n$-even
modes (normal parity) and will couple longitudinally
to all $n$-odd modes with abnormal parity. 
This fixes $z = 1$, hence
$f_\pi = 2/R{g}$, and the longitudinal coupling
becomes $(gf_\pi/\sqrt{2}) \Tr(\partial_\mu\tilde{\pi} A^{1\mu})$.
We emphasize, however, that the results for the 
WZW term will be independent of the choice
of a particular $h(y)$ provided $h(0)=0$ and $h(R)=1$.

An astute reader may be concerned at this point that the 
$\Tr(G_{\mu 5})^2 \sim \Tr (d\tilde{\pi})^2$ 
kinetic term has the form of a linear
realization of chiral symmetry, and does not represent the nonlinear
form embodied in the Wilson line, $W$.  
In fact, if we discard the $\Tr(G_{\mu 5})^2 $ kinetic term
it will not be regenerated by fermion loops, and in this regard can
be considered somewhat unnatural.  What will be generated by fermion
loops is a kinetic term built of 
$W\equiv U \equiv \exp(2i\tilde{\pi}/f_\pi) $, 
and an effective redefinition
of $G_{\mu5}$ given by:
\bea
&& {G}_{\mu 5}\rightarrow \tilde{G}_{\mu 5}  =  
\tilde{U}^\dagger [D_{\mu}, U]  
\nonumber \\
 &&  -\frac{1}{\widetilde{g}^2}\int_0^R dy\int d^4 x
\Tr G_{\mu 5}G^{\mu 5} 
 \rightarrow  
\frac{1}{{g}^2}\int d^4 x
\Tr [D_\mu, {U}][D^\mu, {U}^\dagger] \; .
\eea
where:
\beq
[D_{\mu}, U] = \partial_\mu U - i A_L(x^\mu) U + i U A_R(x^\mu)
\eeq
We see that we have now obtained
a familiar gauged nonlinear $\sigma$-model.
This is what
we would write directly 
in a latticized extra dimension \cite{hill3,zachos}. It
is an amusing exercise to write $\tilde{G}_{\mu 5}$ 
in the continuum theory as a power series
of operators, containing powers of $\partial_y$. 

An important comment is in order presently, which
anticipates the subsequent analysis. The gauge transformation,
$V(y)$, and redefinition, $\tilde{U}(y)$, 
operationally appear to break parity, \ie, the $L\leftrightarrow R$ 
symmetry of the $D=5$ theory, since they each
start on a preferred brane, $I$, and run into the bulk toward $II$. 
However, this asymmetric choice does not, of course, 
{\em physically} break parity, and it has an added bonus:
in the following derivation of the WZW term we find that the CS term
(the ``anomaly flux'' term),
under $\tilde{U}$, develops a parity asymmetric form (\eg, it will
contain terms like $A_R (U^\dagger dU)^3$,
where $U=\tilde{U}(R)$, but no 
corresponding parity conjugates like $A_L (U d U^\dagger)^3$, \etc).
The boundary term (the ``anomaly return flux'') will
likewise develop a parity asymmetric form, and will contain
the parity conjugates, (\eg, the $A_L (Ud U^\dagger )^3$ term
and no $A_R (U^\dagger dU)^3$). In this way, upon
adding the boundary and the CS term,  the overall 
parity symmetry is maintained. The asymmetric choice
of $\tilde{U}$ provides a check on the detailed calculation since
the parity counterparts are split between the two separate 
terms, but sum to the parity symmetric final expression.
It also reveals the roles of the various components of the WZW
term in the masssless fermion case studied in Section IV.

\subsection{The $D=5$ Yang-Mills Chern-Simons Term}

In Appendix A we give some general background information
on the $D=5$ Chern-Simons term. We also show in detail how the 
consistent anomaly matching
to the quarks on the boundary branes leads to quantization
of the CS coefficient. 

In the original fields, the $D=5$ CS term takes the form
(see Appendix A):
\bea
\label{CSterm0}
{S}_{CS} & = & c\int d^5x\; \epsilon^{ABCDE}
\Tr\Bigl ( B_A \partial_B B_C \partial_D B_E   
 - \frac{3i}{2}B_A B_B B_C \partial_D B_E 
- \frac{3}{5}B_A B_B B_C B_D B_E) \Bigr )
\nonumber \\
\eea
where we show that the cancellation of the CS anomalies with
the fermion anomalies on the boundaries requires:
\beq
c = \frac{N_c}{24\pi^2}.
\eeq
Since we are compactifying $x^5$, it is important to
write the CS term in a form
that separates the $B_5$ and $\partial_5$ terms. We obtain
 \cite{zachos}:
\bea
\label{CS11}
S_{CS} = 
 \frac{c}{2}\Tr\int d^4x \int_0^Rdy\;\Bigl[ (\partial_5B_\mu)K^\mu
+\frac{3}{2}\epsilon^{\mu\nu\rho\sigma}\Tr(B_5 
G_{\mu\nu}G_{\rho\sigma})\Bigr],
\eea
where we find:
\bea
K^{\mu} \equiv  \epsilon^{\mu\nu\rho\sigma}
\left( iB_\nu B_\rho B_\sigma + G_{\nu\rho}B_\sigma
+ B_\nu G_{\rho\sigma} \right).
 \eea
In deriving this result, some irrelevant total  divergences in
the $D=4$ subspace have been discarded.
 
Now, performing the transformation of eq.(\ref{trans}),
leading to axial gauge in which $\tilde{B}_5 = 0$,  we see that
the Chern-Simons term becomes:
\bea
\label{CS2}
S_{CS} & = &  
 \frac{c}{2}\epsilon^{\mu\nu\rho\sigma}\int d^4x \int_0^R dy\;
 \Tr \left[\partial_y \tilde{B}_\mu\left(i\tilde{B}_\nu \tilde{B}_\rho \tilde{B}_\sigma 
 + G(\tilde{B})_{\nu\rho}\tilde{B}_\sigma
+ \tilde{B}_\nu G(\tilde{B})_{\rho\sigma}\right)\right] 
\eea
This is the desired form for $S_{CS}$.

 Form notation will be
used throughout the following derivation. This
amounts to simply suppressing indices 
and $\epsilon_{\mu\nu\rho\sigma}$,
\eg, rewriting eq.(\ref{CS2}) using forms is trivial:
\bea
\label{CSf}
S_{CS} & = & \frac{c}{2}\Tr\int d^4 x \int_0^R dy \;
(\partial_y \tilde{B})(2d\tilde{B} \tilde{B} + 2\tilde{B} d\tilde{B} 
- 3i \tilde{B}^3)
\eea
where the Yang-Mills field strength 2-form is: 
$G(\tilde{B})=2d\tilde{B}-2i\tilde{B}^2$.

\subsection{The Boundary Term}

In the large fermion mass limit, integrating
out the fermions,  the Dirac determinant yields an
effective action, $S_{boundary}$.  This
is the  ``boundary term,'' or ``anomaly flux return,'' 
and it arises directly from triangle and box loops of
the fermions with external gauge fields on the boundaries. 
We have explicitly computed
this in the QED case in \cite{hill,hill2} where we find
that the boundary term  is 
equivalent to the (negative of) Bardeen's counterterm, 
$\sim (6\pi^2)^{-1}AVdV$. This counterterm
adds to the pure fermionic action and
transforms the consistent 
anomalies into covariant ones.  We assume that
this result generalizes to the case of Yang-Mills and thus
postulate
the form of the boundary term to be the Yang-Mills
generalization. In fact,
we will see that this form of the boundary term is required to
maintain the full parity invariance of the resulting
WZW term, using the asymmetric
$\tilde{U}(y)$.

 We expect, therefore, in the large $m$ limit
and integrating out the fermions, we will obtain the 
effective boundary action amongst gauge fields of the form:
\bea
\label{boundary}
\label{cobrane}
S_{boundary} & = & -\frac{c}{2}
\int \Tr\big(\half(G_R\tilde{B}_R + \tilde{B}_RG_R)\tilde{B}_L 
-\half(G_L\tilde{B}_L + \tilde{B}_LG_L)\tilde{B}_R  
\nonumber \\ & &
\qquad +i\tilde{B}_R^3\tilde{B}_L - i\tilde{B}_L^3\tilde{B}_R 
-\frac{i}{2}(\tilde{B}_R\tilde{B}_L)^2\big)
\eea
where $G_X = G(\tilde{B}_X) = 2d\tilde{B}_X-i\tilde{B}_X^2$.
Thus $\tilde{S}$ is the sum of eq.(\ref{boundary}) and
eq.(\ref{CS2}).

\section{Derivation of the Wess-Zumino-Witten Term}

Our procedure is now to insert the $\tilde{B}$ field,
written in terms of $A$ and $\tilde{U}$,
into $S_{CS}$ and $S_{boundary}$. We develop the $S_{CS}$ 
maximally into into exact differentials. The sum 
$S_{CS} + S_{boundary}$ yields the WZW term in the large $m$ limit.

As a short-hand notation in what follows, the gauge transformed vector
potentials can be written as:
\beq
\label{pots}
\tilde{B}_\mu = \tilde{A}_\mu - i\alpha_\mu  
\eeq
where:
\beq
\alpha_\mu = -\tilde{U} \partial_\mu \tilde{U}^\dagger \qquad \qquad
 \tilde{A}_\mu = \tilde{U} A_\mu \tilde{U}^\dagger
\eeq
where $\tilde{U}(y)$ is defined in eq.(\ref{uy}).
Note that both $\alpha$ and $\tilde{A}$ are functions of
$x_\mu$ and $y$, and functionals of $\tilde{\pi}$.
At the special values of $y=0$ and $y=R$
we have:
\beq
U \equiv \tilde{U}(R), \quad\qquad
\alpha_{L\mu} = 0\; , \qquad \qquad
\alpha_{R\mu} = \alpha_\mu(x_\mu,R) = -{U} \partial_\mu {U}^\dagger\; ,
\eeq
and we have:
\beq
\tilde{A}_{L\mu} = A_{L\mu} = A_{\mu}(x^\mu, 0)\;, \qquad \qquad
\tilde{A}_{R\mu} = U A_{\mu}(x^\mu, R)U^\dagger .
\eeq
Note again the consequence of the parity asymmetry of $\tilde{U}$
which causes $\alpha_{L\mu} =0$.  For future reference we also
define the conjugate chiral current:
\beq
\beta_{\mu} = U^\dagger \partial_\mu U = U^\dagger \alpha_\mu U
\eeq 
Our notation is identical to that of \cite{kay} for the
purpose of easy comparison (note that
the $\alpha$ and $\beta$ are {\em  not} the usual chiral currents
constructed out of $\xi$, where $\xi^2 = U$, See section IV).
We have the following lemmas:
\bea
d\alpha & = & -dU dU^\dagger = \alpha^2 \qquad\qquad d\beta= -\beta^2
\nonumber \\
dU & = & -U dU^\dagger U = \alpha U  \qquad\qquad
dU^\dagger  =  -U^\dagger dU U^\dagger = -\beta U^\dagger
\eea

\subsection{The Chern-Simons Term}

We'll consider $S_{CS}$ and
$S_{boundary}$ separately.
We first substitute $\tilde{B} = \tilde{A}-i\alpha $ into $S_{CS}$
of eq.(\ref{CSf}):
\bea
S_{CS} 
& = & \frac{c}{2}\Tr\int d^4 x\; dy \; 
[-i(\partial_y \alpha) + (\partial_y \tilde{A})]
\nonumber \\
& & \times  \big( 
2d\tilde{A}\tilde{A} - 2i\alpha^2\tilde{A} 
-  2id\tilde{A}\alpha - 4\alpha^3 
+ 2\tilde{A}d\tilde{A}- 2i\tilde{A}\alpha^2 
- 2i\alpha d\tilde{A} 
\nonumber \\
& & \qquad \qquad  - 3i \tilde{A}^3 -3 \alpha \tilde{A}^2 
-3 \tilde{A}\alpha\tilde{A} -3\tilde{A}^2\alpha 
+3i \alpha^2 \tilde{A} + 3i\alpha \tilde{A}\alpha +3i \tilde{A}\alpha^2 
+ 3\alpha^3) \nonumber \\
\eea

The trick presently is to write this expression
in terms of exact-differentials in $y$, which leads to
exact integrals over $y$.  It is convenient at present to set $R=1$
and treat $y$ as a dimensionless integration variable running from
$0$ to $1$. We also use ``$\int$'' to represent 
``$\int d^4x\;\int_0^1 dy$'' in the following, unless otherwise specified.
The analysis is straightforward, but a casual reader may skip to
the results for $S_{CS}$, given below in
eq.(\ref{resultcs}). The details of the derivation follow presently. 

\subsubsection{The Original Wess-Zumino Term: $(\partial_y\alpha) \alpha^3$ }

We first isolate the term:
\bea
S_{CS 0} & = & i\frac{c}{2}\Tr\int (\partial_y \alpha)  \alpha^3
\eea
With $\tilde{U}=\exp(2i\tilde{\pi}y/f_\pi )$, 
we note the {\em exact} result:
\beq
\partial_y \alpha = \partial_y \tilde{U} d \tilde{U}^\dagger 
= \frac{2i}{Rf_\pi}\tilde{U}
d\tilde{\pi} \tilde{U}^\dagger  
\eeq
If we expand the remaining $\alpha \approx 2iyd\tilde{\pi}/f_\pi$ then
we obtain a null result $\propto 
\Tr(d\tilde{\pi}d\tilde{\pi}d\tilde{\pi}d\tilde{\pi})$, vanishing by
cyclicity of the trace. However, to the next order in
expansion (consistently for all $\alpha$ factors), we obtain:
\beq
\alpha \approx  \frac{2iy}{f_\pi}d\tilde{\pi} -  \frac{2y^2}{f^2_\pi}
[\tilde{\pi}, d\tilde{\pi}] + ...
\eeq
Thus, we find, using $c=N_c/24\pi^2$:
\bea
S_{CS0} & = & -\frac{2N_c }{3\pi^2 f_\pi^5}
\int d^4x \; dy y^4 \Tr (\tilde{\pi}d\tilde{\pi}d\tilde{\pi}d\tilde{\pi}d\tilde{\pi})
+ ...
\nonumber \\
& = & -\frac{2N_c }{15\pi^2 f_\pi^5}
\int d^4x  \Tr ( \tilde{\pi}d\tilde{\pi}d\tilde{\pi}d\tilde{\pi}d\tilde{\pi})
+ ...
\eea
This is the original Wess-Zumino term with Witten's quantized
coefficient.

\subsubsection{The $\alpha^3 \tilde{A}$ Term}

We now collect together terms of the form:
\bea
S_{CS\;\alpha^3 \tilde{A}} & = & -i\frac{c}{2}\Tr\int  (\partial_y \alpha)
\bigl(-2id\tilde{A}\alpha - 2i\alpha d\tilde{A}
- 2i\alpha^2\tilde{A}- 2i\tilde{A}\alpha^2 
+3i (\alpha^2 \tilde{A} +\alpha \tilde{A}\alpha + \tilde{A}\alpha^2) 
 \bigr)\nonumber \\
& & 
-\frac{c}{2} \Tr\int
(\partial_y \tilde{A}) [\alpha^3 ]
\eea
Note that, upon integrating in $D=4$ by parts:
\beq
\Tr\int  (\partial_y \alpha)
\big(d\tilde{A}\alpha +\alpha d\tilde{A})
= 2\Tr\int  (\partial_y \alpha)
\big(\alpha \tilde{A}\alpha )
\eeq
Thus, we can immediately write:
\bea
S_{CS\;\alpha^3 \tilde{A}} & = & -i\frac{c}{2}\Tr\int (\partial_y \alpha)
\big( i\alpha^2\tilde{A}+i\tilde{A}\alpha^2 
-i\alpha \tilde{A}\alpha  \big)
-\frac{c}{2}\Tr\int d^4 x dy (\partial_y \tilde{A})[ \alpha^3 ]\nonumber \\
& = & \frac{c}{2}\Tr\int d^4x \int_0^1 dy\; \partial_y (\alpha^3 \tilde{A})
\eea
If we now explicitly perform this integral we obtain:
\bea
S_{CS\;\alpha^3 \tilde{A}} & = & -\frac{c}{2}\Tr({A}_R\beta^3 ) 									   
\eea
where use has been made $\Tr(\alpha^3 \tilde{A}_R)  = 
\Tr(\alpha^3 U{A}_R U^\dagger ) = \Tr(U^\dagger \alpha^3 U{A}_R  )
=\Tr(\beta^3 A_R) = -\Tr( A_R\beta^3)$.
We see the operational parity asymmetry of our gauge tranformation 
leads to the absence of a corresponding parity conjugate
term, $-\Tr( A_L\alpha^3)$. As mentioned above, this term will come
from the boundary term, and the overall final result will be
parity symmetric.

\subsubsection{The $\alpha \tilde{A}^3$ Term}

We now collect terms of the form:
\bea
S_{CS\;\alpha \tilde{A}^3} +S^r & = & -i\frac{c}{2}\Tr\int  (\partial_y \alpha)
\big( 2d\tilde{A} \tilde{A} + 2\tilde{A}d\tilde{A} -3i\tilde{A}^3\big) 
\nonumber \\
& & 
+\frac{c}{2} \Tr \int  (\partial_y \tilde{A}) [2d\tilde{A}\tilde{A}+2\tilde{A}d\tilde{A}
-3(\alpha\tilde{A}^2 +\tilde{A}\alpha\tilde{A} + \tilde{A}^2\alpha)]
\eea
where $S^r$ is a remainder (see below).
We now use 
$d\tilde{A} = \alpha\tilde{A} +\tilde{A}\alpha +U d A U^\dagger$,
and the lemma:
\beq
\Tr\int  \partial_y (\alpha\tilde{A}^3)=
\Tr\int  [ (\partial_y \alpha)(\tilde{A}^3)-
(\partial_y \tilde{A})(\tilde{A}^2\alpha -\tilde{A}\alpha\tilde{A} 
+ \alpha\tilde{A}^2)]
\eeq
to write:
\bea
\label{aA3}
S_{CS\;\alpha \tilde{A}^3} & = & 
+\frac{c}{2}\Tr\int \partial_y (\alpha\tilde{A}^3)
\eea
and the remainder:
\bea
\label{aA3r}
S^r & = & -i\frac{c}{2}\Tr\int  (\partial_y \alpha)
\big( 2d\tilde{A} \tilde{A} + 2\tilde{A}d\tilde{A} -4i\tilde{A}^3\big) 
+\frac{c}{2}\Tr\int  (\partial_y \tilde{A}) [2\tilde{U}(d{A}{A}+{A}d{A})
\tilde{U}^\dagger ]
\eea
The remainder
is carried into the next set of $\alpha^2 \tilde{A}^2$ terms.

\subsubsection{The $\alpha^2 \tilde{A}^2$ Terms}

Including the remainder from eq.(\ref{aA3r}), we now have
the residual terms:
\bea
S_{CS\;\alpha^2 \tilde{A}^2}& = & 
-i\frac{c}{2}\Tr\int  (\partial_y \alpha)
\big( 2d\tilde{A} \tilde{A} + 2\tilde{A}d\tilde{A} -4i\tilde{A}^3 
-3 \alpha \tilde{A}^2 
-3 \tilde{A}\alpha\tilde{A} -3\tilde{A}^2\alpha \big) \nonumber \\
& &
+ \frac{c}{2}\Tr\int  (\partial_y \tilde{A})\big(
[\tilde{U} (2d{A}{A}+2{A}d{A})\tilde{U}^\dagger]
+i \alpha^2 \tilde{A} 
\nonumber \\ & & \qquad 
+ 3i\alpha \tilde{A}\alpha +i \tilde{A}\alpha^2 
-2id\tilde{A}\alpha - 2i\alpha d\tilde{A} - 3i\tilde{A}^3)
\eea
Using:
\bea
-\Tr(\partial_y(d\tilde{A}\tilde{A}\alpha) & = &
\Tr((\partial_y\alpha) d\tilde{A}\tilde{A})
-\Tr((\partial_y \tilde{A})(d\tilde{A}\alpha 
-\tilde{A}\alpha^2+\alpha d\tilde{A} ) )
\nonumber \\
\Tr(\partial_y(\alpha\tilde{A}d\tilde{A}) & = &
\Tr((\partial_y\alpha) \tilde{A}d\tilde{A})
-\Tr((\partial_y \tilde{A})(d\tilde{A}\alpha 
-\alpha^2\tilde{A}+\alpha d \tilde{A} ) )
\nonumber \\
\Tr(\partial_y(\alpha \tilde{A}\alpha\tilde{A}) & = &
2\Tr((\partial_y\alpha) \tilde{A}\alpha\tilde{A})
-2\Tr((\partial_y \tilde{A})\alpha\tilde{A}\alpha)
\eea
and we therefore have:
\bea
S_{CS\;\alpha^2 \tilde{A}^2}
& = & -i\frac{c}{2}\Tr\int  (\partial_y \alpha)
\big( \tilde{U}( 3d{A} {A} + 3{A}d{A} 
-4i{A}^3 )\tilde{U}^\dagger ) \nonumber \\
& &
+ \frac{c}{2}\Tr\int  (\partial_y \tilde{A})
[\tilde{U} (2d{A}{A}+2{A}d{A}- 3i{A}^3)\tilde{U}^\dagger] 
\nonumber \\
& & 
+ \frac{c}{2}\int   \Tr\bigl[\partial_y (\tilde{U} d{A}\tilde{U}^\dagger
(-i\tilde{A}\alpha
+i\alpha\tilde{A})) \bigr]
+\frac{ic}{4} \int  \Tr[\partial_y(\alpha \tilde{A}\alpha\tilde{A})].
\eea
Note that we have used the identity, 
$\tilde{U}dA \tilde{U}^\dagger 
= d\tilde{A} - \alpha \tilde{A} - \tilde{A} \alpha $,
to remove the $\tilde{} $
from the $\tilde{A}$ fields that sandwiched 
between $\tilde{U}$ and $\tilde{U}^\dagger$
in the above expression.

\subsection{Summary of Results for $S_{CS}$}

Collecting the results $S_{CS\;\alpha \tilde{A}^3}$, 
$S_{CS\;\alpha \tilde{A}^3}$ 
and $S_{CS\;\alpha^2 \tilde{A}^2}$ and
performing the exact integrals, we have:
\bea
\label{resultcs}
S_{CS} & = & S_{CS0}-\frac{c}{2}\Tr(A_R\beta^3 ) -\frac{c}{2}\Tr(A^3_R\beta)
-i\frac{c}{4}   \Tr( {A}_R\beta {A}_R\beta )
- i\frac{c}{2}\Tr\bigl[(d{A_R}{A}_R+{A}_Rd{A_R})\beta  \bigr]
\nonumber \\
& & 
-i\frac{c}{2}\Tr\int  (\partial_y \alpha)
\big( \tilde{U}( 3d{A} {A} + 3{A}d{A} 
-4i{A}^3 )\tilde{U}^\dagger ) 
\nonumber \\
& &
+ \frac{c}{2}\Tr\int  (\partial_y \tilde{A})
[\tilde{U}(2d{A}{A}+2{A}d{A}- 3i{A}^3)\tilde{U}^\dagger ] 
\eea
This is the pure Chern-Simons action. As mentioned
above, it is
parity asymmetric owing to the asymmetric definition of
$\tilde{U}(y)$, and the boundary term will restore the parity symmetry. 

We now compute the boundary term (anomaly flux return). 
The final results
are quoted in eq.(\ref{final}).

\subsection{Boundary Term (Anomaly Flux Return)}

We substitute eq.(\ref{pots}) into eq.(\ref{cobrane})
and straightforwardly evaluate.
We note that $G_R (\tilde{B}_R) \rightarrow U G_R(A_R) U^\dagger$,
and $G_L(\tilde{B}_L)\rightarrow G_L(A_L)$.  The result is:
\bea
S_{boundary} & = & 
 \frac{c}{2}  \int \Tr\big[  
 (dA_L A_L + A_L dA_L )U A_R U^\dagger -
 (dA_R A_R + A_R d A_R) U^\dagger A_L U
 \nonumber \\
 & & -i(dA_LA_L + A_L dA_L)\alpha -A_L^3\alpha-A_L\alpha^3
 +iA_R^3 U^\dagger A_L U -iA_L^3 U A_R U^\dagger
\nonumber \\
& & -i(dA_R dU^\dagger A_L U  -dA_L dU A_R U^\dagger)  
- (A_R U^\dagger A_L U A_R \beta   + A_L U  A_R U^\dagger A_L \alpha) 
\nonumber \\
&& 
+ \frac{i}{2}  A_L \alpha A_L \alpha
+ \frac{i}{2} U  A_R U^\dagger A_L U  A_R U^\dagger A_L   
-i(A_L U {A}_R U^\dagger \alpha^2 - {A}_R U^\dagger A_L U \beta^2)
\big]
\nonumber \\
\eea
where we have used:
\bea
-i(A_L U dA_RU^\dagger + U^\dagger dA_R U A_L)\alpha 
+i(U \beta A_R U^\dagger)\alpha A_L
 = -i(dA_R dU^\dagger A_L U 
-dA_L dU A_R U^\dagger)  
\nonumber \\
\eea
We see, as in the case of the Chern-Simons anomaly flux, that
the result is parity asymmetric, a consequence of our
asymmetric choice of $U(y)$. However, we now recover a fully 
parity symmetric form when we combine the CS term and boundary terms.


\subsection{Full Wess-Zumino-Witten Term}

If we now combine all terms, we have the full Wess-Zumino-Witten term,
derived from $\tilde{S}=S_{CS} + S_{boundary}$ of eq.(\ref{tilS}),
and where we now define:
\beq
\tilde{S} =  S_{WZW} + S_{bulk} 
\eeq
where:
\bea
\label{final}
{S}_{WZW} & = & S_{CS0}+\frac{N_c}{48\pi^2}\Tr\int d^4x \bigl[ 
-({A}_L\alpha^3 + {A}_R\beta^3 ) -({A}^3_L \alpha+{A}^3_R \beta)
\nonumber \\
& & 
-i ((dA_LA_L + A_L dA_L)\alpha+d{A}_R{A}_R +{A}_RdA_R)\beta)
+\frac{i}{2}[({A}_L\alpha)^2 - ({A}_R\beta)^2]
\nonumber \\
& &
-i(A_L^3 U  A_R U^\dagger - A_R^3 U^\dagger A_LU) 
\nonumber \\
& &
+ (dA_L A_L + A_L dA_L )U A_R U^\dagger -(dA_R A_R + A_R d A_R) U^\dagger A_L U
\nonumber \\
& & 
-i ( dA_R dU^\dagger A_L U - dA_L dU A_R U^\dagger  )
-(A_L U  A_R U^\dagger A_L \alpha + A_R U^\dagger A_L U A_R \beta  ) 
\nonumber \\
& & 
+ \frac{i}{2} U  A_R U^\dagger A_L U  A_R  U^\dagger A_L   
-i(A_L U {A}_R U^\dagger \alpha^2 - {A}_R U^\dagger A_L U \beta^2)\bigr]
\nonumber \\ 
\tilde{S}_{CS0} & = & -\frac{2N_c }{15\pi^2 f_\pi^5}
\int d^4x  \Tr ( \tilde{\pi}d\tilde{\pi}d\tilde{\pi}d\tilde{\pi}d\tilde{\pi})
+ ... 
\eea
$S_{WZW}$ is
seen to be in complete agreement with Kaymakcalan, Rajeev and Schechter
\cite{kay} (our result differs by
an overall minus sign).

The remaining term, $S_{bulk}$, is built of inexact integrals over the
bulk ($c=N_c/24\pi^2$):
\bea
S_{bulk} & = & 
-i\frac{c}{2}\Tr\int  (\partial_y \alpha)
\big( \tilde{U}( 3d{A} {A} + 3{A}d{A} 
-4i{A}^3 )\tilde{U}^\dagger ) \nonumber \\
& &
+ \frac{c}{2}\Tr\int  (\partial_y \tilde{A})
[\tilde{U}(2d{A}{A}+2{A}d{A}- 3i{A}^3)\tilde{U}^\dagger ] 
\eea 
These are readily developed using the exact results:
\beq
\partial_y \alpha = \frac{2i}{f_\pi} \tilde{U} (d\tilde{\pi}) \tilde{U}^\dagger
\qquad\qquad
\partial_y\tilde{A} =\partial_y\tilde{U}{A}\tilde{U}^\dagger
= \frac{2i}{f_\pi} \tilde{U} ([\tilde{\pi},A]) \tilde{U}^\dagger
\eeq
Note that these expressions are valid to all orders
in $\tilde{\pi}$ (not truncated expansions in
the $\tilde{\pi}$).
Substituting, we see that:
\bea
S_{bulk} & = & 
-\frac{3c}{2f_\pi}\int d^4x \int_0^1 dy \Tr(\tilde{\pi} G G)
+
 \frac{c}{2}\int d^4x \int_0^1 dy \Tr
  (\partial_y {A})(2d{A}{A}+2{A}d{A}- 3i{A}^3)) 
  \nonumber \\
\eea 
We see by comparison
to eq.(\ref{CS11}) with the matching $A_5 = -2\tilde{\pi}/f_\pi$
that $S_{bulk}$ is just the Chern-Simons term written in the 
new field variables, $A_\mu$ and $\tilde{\pi}$. 
This reflects bulk interactions amongst KK-modes. 
One can obtain the detailed form of these
interactions by substituting the wave-functions in the bulk for the KK-modes
and performing the $dy$ integrations, as was done previously
for QED \cite{hill}.  The sum of the bulk and boundary contributions 
make this physics gauge invariant.

Indeed, under a gauge transformation, the WZW term
yields the (negative of the) consistent
anomaly \cite{kay}. Likewise, the bulk term
yields the  consistent anomaly, and taken
together, these contributions cancel.
This happens because we have started with a gauge invariant
theory.  However, a $D=4$ chiral lagrangian of mesons
has no bulk interaction term, and it is anomalous.
The result for such a theory is just the $S_{WZW}$ term alone,
as is well known.

Eq.(\ref{final}) is the general result for any $D=5$
system involving chiral delocalization and bulk Yang-Mills fields.
We need only substitute $A_\mu(x,y) = \sum A_\mu^n(x,y)$ and
$A_{L\mu}(x) = A_\mu(x,0)$, and $A_{R\mu}(x) = A_\mu(x,R)$,
and identify $2\tilde{\pi}/f_\pi =-\int dy A_5$.

\section{Massless Fermions}

We can now do something novel with this formalism.
It is useful
to consider the form of $\tilde{S}$ when the fermions
have a small mass and are not integrated out.
We envision many possible applications
in this limit, since many modern theories are effectively extra dimensional
with chiral delocalization. For example, Little Higgs bosons
are essentially PNGB's, similar to $K$-mesons, and
the fermion content of these models is effectively a
chirally delocalized system in $D=5$ (usually described by
a form of deconstruction). This form of the WZW term would be applicable
to Little Higgs interactions with other PNGB's in the theory.
For example, we would expect $H + H^\dagger \rightarrow 3\tilde{\pi}$
proceeding though the $\Tr(\pi (d\pi)^4) $ term. 

First, it is useful to write the parity asymmetric form,
which follows directly from the results derived above.
If the fermion mass $m$ is small, and the fermions unintegrated, then
the boundary term is not present, but the $S_{CS}$ will be.   
We can immediately write
the form of the effective lagrangian from eq({\ref{resultcs}):  
\bea
\label{resultcs2}
S & = & S_{CS0}-\frac{c}{2}\Tr(A_R\beta^3 ) -\frac{c}{2}\Tr(A^3_R\beta)
-i\frac{c}{4}   \Tr( {A}_R\beta {A}_R\beta )
- i\frac{c}{2}\Tr\bigl[(d{A_R}{A}_R+{A}_Rd{A_R})\beta  \bigr]
\nonumber \\
& &
+\int_I d^4x \;\bar{\psi}_L(i\slash{\partial}+\slash{A}_L)\psi_L 
+  \int_{II} d^4x \;\bar{\psi}_R(i\slash{\partial}+
U(\slash{A}_R  - i\beta)U^\dagger) \psi_R
+S_{bulk}
\eea
This form is revealing.  The theory is fully gauge invariant, and 
we thus see that a
gauge transformation on  $A_L$ commutes with $\beta$. Hence, only
the $S_{bulk}$ shifts, producing the consistent anomaly
that cancels the fermionic anomaly. On the other hand,
we can view $U(\slash{A}_R  - i\beta)U^\dagger$ as a Stueckelberg
field, and a shift of $\delta A_R = d\theta_R$ is compensated by 
$\delta \beta= -d\theta_R$ (recall that $A_5= - 2i\tilde{\pi}/f_\pi$),
and no fermionic anomaly is generated.
The theory must be invariant under this tranformation,
and we see that this happens by a cancellation between $S_{bulk} $ and
the first four terms of eq.(\ref{resultcs2}).  This provides
a shorthand derivation of the fact that, under a gauge
transformation,  $S_{bulk}$ cancels
$S_{WZW}$ in eq.(\ref{final}).

We can cast the above results into a form that is parity symmetric.
We redefine $\tilde{U}(y)$ as ($R=1$):
\beq
\tilde{U}(y) = \exp\left(\frac{2i\tilde{\pi}(y-1/2)}{f_\pi }\right)
\eeq
We can define:
\beq
U(R) = \xi \qquad \qquad U(0) =\xi^\dagger 
\eeq
The current $\alpha(y) = -U(y)dU^\dagger(y)$, 
$\tilde{B} =\tilde{A} -i\alpha$ and $\tilde{A}= U(y)A U^\dagger $
are as defined previously, but now
we have:
\beq
\tilde{B}_L = \xi A_L \xi^\dagger -j_L
\qquad \qquad
\tilde{B}_R = \xi^\dagger  A_L \xi-j_R
\eeq
where:
\beq
j_L = i\xi d \xi^\dagger
\qquad \qquad 
j_R = -i\xi^\dagger d\xi
\eeq
and $S_{CS}$, added to the fermionic action, 
thus becomes, from eq(\ref{resultcs}):
\bea
\label{resultcs2}
S & = & S_{CS0} + S'_{WZW}+S_{bulk}
\nonumber \\
& &
+\int_I d^4x \;\bar{\psi}_L(i\slash{\partial}+
\xi\slash{A}_L \xi^\dagger  -j_L)\psi_L 
+  \int_{II} d^4x \;\bar{\psi}_R(i\slash{\partial}+\xi^\dagger\slash{A}_R\xi
-j_R)\psi_R
\eea
where:
\bea
 S'_{WZW} & = & -\frac{c}{2}\Tr(A_Rj_R^3 + A_Lj_L^3) 
-\frac{c}{2}\Tr(A^3_Rj_R+A^3_Lj_L)
-i\frac{c}{4}   \Tr( {A}_Rj_R {A}_Rj_R - {A}_Lj_L {A}_Lj_L )
\nonumber \\
& & 
- i\frac{c}{2}\Tr\bigl[(d{A_R}{A}_R+{A}_Rd{A_R})j_R
+ (d{A_L}{A}_L+{A}_Ld{A_L})j_L \bigr]
\eea
Note that $S_{CS0}$ and
$S_{bulk}$ are unchanged in form.
$S'_{WZW}+S_{bulk}$ generates the consistent anomalies
to cancel the fermionic anomalies under the various
forms of gauge and local chiral tranformations. 

\newpage
\section{Conclusions}

We have shown that the Chern-Simons term of a $D=5$ Yang-Mills theory,
together with the boundary terms, yields the full Wess-Zumino-Witten
term of a $D=4$ gauged chiral lagrangian.  
The present analysis was
possible after insights were gleaned from earlier work \cite{hill}
and \cite{hill2}, which considered in detail the $U(1)$ theory
(QED) in $D=5$ with chiral electrons on boundary branes.
In yet another earlier paper we developed
the relevant form of the CS-term under compactification
of $x^5$, and we attempted to construct the full  WZW term from
a pure Yang-Mills theory using latticization
(deconstruction) \cite{zachos}. This approach did not yield the
full gauge structure, which we have achieved presently.
The present analysis is essentially a detailed 
application of \cite{hill} to Yang-Mills theories.

Let us summarize how the analysis proceeds in general.  We begin 
in a $D=5$ Yang-Mills theory, compactified
in $0 \leq x^5 \leq R$, with chirally delocalized fermions
on the boundaries (branes).
The theory contains a bulk-filling Chern-Simons term.
The chiral fermions have a gauge invariant mass term that is bilocal,
$\sim\bar{\psi}_L(x,0)W\psi_R(x,R)+h.c.$, and
involves the Wilson line, $W=P\exp(i\int_0^R B_5dx^5)$ 
that spans the bulk. The Wilson line is identified
with a chiral field of mesons, $W=\exp(2i\tilde{\pi}/f_\pi)$.
A general gauge transformation in the bulk produces anomalies on the
boundaries coming from the Chern-Simons term.  
Likewise, this gauge transformation produces anomalies, coming
from the fermions on the boundaries. These anomalies take the consistent form,
\ie, they are the direct result of the Feynman triangle loops for the
fermions, and have the identical form 
as the anomalies from the CS term (see Appendix).
We demand that these anomalies cancel, and this fixes the coefficient of the 
CS term, generally to $c=N_c/24\pi^2$.

We now rewrite the CS term into a form that displays separately $B_5$ 
and $\partial_5$. We then perform a master gauge transformation that converts
$B_5\rightarrow 0$. This also sets the Wilson line spanning the bulk between
the branes to unity. This results in a field $\tilde{B} $ that has the
mesons comingled with gauge fields. We thus redefine
$\tilde{B} = \tilde{U}{A}\tilde{U}^\dagger + \alpha$, where  $\tilde{U}(y) = 
\exp(2iy\tilde{\pi}/Rf_\pi)$, and $\alpha =-\tilde{U} d \tilde{U}^\dagger$
is a chiral current built of the mesons. 
This separates the $\tilde{\pi}$ mesons from
the physical gauge fields $A$. Moreover, the massive components of $A$ are
now gauge covariant Stueckelberg fields (see \cite{hill}), having ``eaten''
their longitudinal degrees of freedom contained 
in the non-zero modes of $B_5$. 

Finally, we integrate out the fermions in the large $m$ limit. This
produces effective interactions (the log of the Dirac determinant) 
on the boundaries. The form of
this effective ``Boundary Term'' interaction is just 
Bardeen's counterterm \cite{bardeen}
that maps consistent anomalies into covariant ones.  We thus have an
expression for total action, $\tilde{S}$, the sum of $S_{CS}$, 
the Chern-Simons term,
and  $S_{boundary}$, the boundary terms from the fermionic Dirac determinant.
These are functionals of the field $\tilde{B} = \tilde{A} -i\alpha$

We now straightforwardly manipulate the $\tilde{S}$ into terms 
that are exact forms in the
$x^5$ dimensions, and produce exact integrals,
yielding terms that depend only upon the fields on the 
boundaries. 
The result is the full Wess-Zumino-Witten term, 
together with bulk interactions amongst KK-modes mediated by the Chern-Simons
term.

We have also given a novel form of the WZW term in the case
that the fermions are not integrated out. This reveals the
roles of the various components of the full WZW term under the various gauge
interactions. 

These results apply, in principle, to any theory with chiral delocalization
in extra dimensions. If all
of the $B_5$ KK modes are eaten, then we can simply set $\tilde{\pi}$
to zero everywhere in eq.(\ref{final}). The remaining terms yield the
gauge invariant physics of new interactions amongst KK-modes that
are generated jointly by the Chern-Simons term and boundary interactions
\cite{hill}.

There are many theories to which these considerations apply, 
but to which, thus far, this essential physics has not
been incorporated. These theories include 
many incarnations
of Randall-Sundrum models, Little Higgs theories, and 
models of (anomaly) split fermion representations
in extra dimensions.  The Little Higgs is a PNGB and
should participate, like the $\pi$ or $K$ mesons of QCD, in 
topological WZW interactions. We further envision applications to
string theory, and AdS-CFT QCD as well (for
a number of related analyses in the context of SUSY
see \cite{gates}, and a similar approach in $M$-theory see
\cite{bilal}). The WZW term of gravitation in
a split anomaly mode, \eg, in $D=6$ and $D=7$, would also be
an intriguing application.   

A more expansive analysis of the 
current algebra associated with the
$D=4$ and $D=5$ chiral/Yang-Mills correspondence 
is underway, and a number of novel applications
is envisioned \cite{hill5}.

Note:  It has been brought to my attention that a previous work of
Sakai and Sugimoto, carried out in the context of string theory
with an ultimately similar configuration to ours, claims to
obtain the WZW term from the $D=5$ CS term \cite{sakai}. 
The authors do not discuss the bulk interaction of our eq.(\ref{final}),
or the unintegrated fermion case of eq.(\ref{resultcs2}). It is unclear
as to how the analogue of the boundary term arises in their analysis.
Nonetheless, their setup and analysis is quite similar to ours in many respects.
It has also been brought to my attention that a little known paper
of Novikov \cite{novikov} first remarked upon the quantization of the coefficient
of the Wess-Zumino term, anticipating the classic
work of Witten \cite{witten}.


\appendix
\section{Structure of $D=5$ Chern-Simons Term }

The $D=5$ Yang-Mills theory of eq.(\ref{kinetic}), 
possesses two conserved 
currents of the form:
\bea
\label{cur1}
J_A & = & \epsilon_{ABCDE}\Tr(G^{BC}G^{DE})  ,
\eea
\bea
\label{cur2}
J^a_A & = & \epsilon_{ABCDE}
\Tr \Bigl (\frac{\lambda^a}{2}\{G^{BC},G^{DE}\}\Bigr ) ~.
\eea
The second current requires that $SU(N)$ possess a $d$-symbol, 
hence $N\geq 3$, and it is covariantly
conserved, $[D^A, ~ J^a_A~\lambda^a/2]=0$. These topological  
currents do not arise from $S_0$ 
under local Noetherian variation of the fields. 

Why do these currents exist? In fact,
these currents describe a special topological soliton in $D=5$, 
the ``instantonic soliton,'' that consists
of an instanton living on an arbitrary time slice \cite{ramond}. Owing
to eq.(\ref{cur1}) the instantonic soliton carries a conserved charge.
Since it is an $SU(2)$ configuration, the current eq.(\ref{cur2})
simply measures how the $SU(2)$ configurations can be imbedded and rotate
within the $SU(N)$ group (hence $d$-symbols measure imbeddings of $SU(2)$
into higher Lie groups). 
When the theory is compactified according to the rules {\em(1)} and {\em (2)},
then this soliton becomes the Skyrmion,  eq.(\ref{cur1}) becomes
the Goldstone-Wilczek current representing baryon number of
the skyrmion, (eq.(\ref{cur2}) becomes a transition flavor
current amongst flavors of baryons). The Chern-Simons term
when added to the Lagrangian becomes the generator
of these currents (see \cite{zachos}), just as the WZW
term is the generator of flavor-skyrmion currents.  These correspondences
are very tight, even at the level of precise mathematical matchings
(\ie, one can infer the form of the full Goldstone-Wilczek current with
gauging by matching to eq.(\ref{cur1}) and using a
latticized compactification).
This correspondence motivates the
search for the correspondence between the full WZW term
and the Chern-Simons term.

The Chern-Simons term (second Chern character) 
takes the form: 
\bea
\label{appCSterm0}
{\cal{L}}_{CS} & = & c\epsilon^{ABCDE}
\Tr\Bigl ( A_A \partial_B A_C \partial_D A_E   
 - \frac{3i}{2}A_A A_BA_C \partial_D A_E 
- \frac{3}{5}A_A A_B A_C A_D A_E)
\eea
and it can be conveniently rewritten as:
\bea
\label{CSterm1}
{\cal{L}}_{CS} & = &
\frac{c}{4}\epsilon^{ABCDE}
\Tr(A_A G_{BC}G_{DE}   
 + i A_A A_B A_C G_{DE}
- \frac{2}{5}A_A A_B A_C A_D A_E \Bigr ) ~.
\eea
It is derived by ascending to $D=6$ and considering the 
generalization of the Pontryagin index
(a $D=6$ generalization of the $\theta$-term),
\beq
{\cal{L}}_P =\epsilon_{ABCDEF}\Tr G^{AB}G^{CD}G^{EF} .
\eeq
which can be written as a total divergence, 
\beq
{\cal{L}}_P 
=
-8\partial^F\epsilon_{ABCDEF}
\Tr \Bigl ( A_A \partial_B A_C \partial_D A_E   
- \frac{3i}{2}A_A A_BA_C \partial_D A_E 
- \frac{3}{5}A_A A_B A_C A_D A_E\Bigr ) . 
\eeq
Formally, compactifying the sixth
dimension and integrating ${\cal{L}}_0$ over 
the boundary in $x^5$ leads to ${\cal{L}}_1$. The Chern-Simons
term can be constructed in any odd
dimension from a general algorithm \cite{Wu}.

Let us perform a generic gauge transformation
in the bulk:
\beq
A_A \rightarrow   V(A_A + i\partial_A)V^\dagger \qquad
\makebox{where:}
\qquad V = \exp(i\theta^a T^a)
\eeq
and we examine the variation of $S_{CS}$ under
this transformation with respect to 
an infinitesimal $\partial_A \theta^a$. 
It is most convenient to
use eq.(\ref{CSterm1}),
since $G_{AB}\rightarrow U^\dagger G_{AB} U$
and we obtain:
\bea
\frac{\delta S_{CS}}{\partial_A\theta^a}
& = &  
c\epsilon^{ABCDE}
\Tr(T^a \partial_B A_C \partial_D A_E )  
\nonumber \\
& &  
-\half i\Tr( T^a A_B A_C (\partial_{D}A_E)
- i T^a  A_B (\partial_{C}A_D) A_E 
+ i T^a (\partial_{B}A_C)A_D A_E )
\eea
If $D=5$ is compactified with boundaries located
at $x^5=0$ and $x^5 = R$, denoted
respectively as $I$ and $II$, 
then under the gauge transformation we have:
\bea
\label{cna}
\delta S_{CS}
& = &  
\left. c\epsilon^{\mu\nu\rho\sigma}\theta^a
\Tr[T^a (\partial_\mu A_\nu \partial_\rho A_\sigma   
-\frac{i}{2}( \partial_\mu A_\nu A_\rho A_\sigma
 - A_\mu \partial_\nu A_\rho A_\sigma + A_\mu A_\nu \partial_\rho  A_\sigma 
) ]\right|^{R}_0
\nonumber \\
\eea
We refer to this as the ``Chern-Simons anomaly.''  

We have introduced chiral quarks on the boundaries $I$ and $II$.
The general gauge transformation $U(x^5) = \exp(iT^a\theta^a (x^\mu, x^5)$ 
acts upon the fermion fields an Wilson line as:
\beq
\psi_L \rightarrow \exp(i\theta (x_\mu, 0))\psi_L \; , \qquad 
\psi_R \rightarrow \exp(i\theta (x_\mu, R))\psi_R \;
\qquad W\rightarrow V(0) W V^\dagger(R)
\eeq 
The fermionic action transforms as:
\bea
\label{shift0}
S_{branes} & \rightarrow  & 
S_{branes}  -
\int_I d^4x \;\theta^a (x_\mu, 0)Y^a_L 
 - \int_{II} d^4x \; \theta^a (x_\mu, R) Y^a_R
\eea
where $Y^a_{L,R}$ is the fermionic anomaly on the corresponding
brane.
We use Bardeen's result for the {\em consistent } nonabelian
anomalies \cite{bardeen}:
\bea
Y_R^a & = &
\frac{1}{24\pi^2}\epsilon^{\mu\nu\rho\sigma} \Tr[T^a (\partial_\mu A_{R\nu} \partial_\rho A_{R\sigma }  
-\frac{i}{2}( \partial_\mu A_{R\nu} A_{R\rho} A_{R\sigma}
 - A_{R\mu} \partial_\nu A_{R\rho} A_{R\sigma} + A_{R\mu} A_{R\nu} 
 \partial_\rho  A_{R\sigma} 
) ] \nonumber \\
Y_L^a & = &
-\frac{1}{24\pi^2} \epsilon^{\mu\nu\rho\sigma}\Tr[T^a (\partial_\mu A_{L\nu} \partial_\rho A_{L\sigma }  
-\frac{i}{2}( \partial_\mu A_{L\nu} A_{L\rho} A_{L\sigma}
 - A_{R\mu} \partial_\nu A_{L\rho} A_{L\sigma} + A_{L\mu} A_{L\nu} 
 \partial_\rho  A_{L\sigma} 
) ]\nonumber \\
\eea
Note that the consistent anomalies are independent
of the mass of the fermion, and they do not decouple
in the $m\rightarrow \infty$ limit (while the
covariant anomalies do decouple).

Thus we see that the CS anomaly has exactly 
the same form as the fermionic consistent anomalies.
This implies that we can cancel the femionic
anomalies against the CS anomalies, if we have:
\beq
c = \frac{N_c}{24\pi^2}\;.
\eeq

\newpage

\vskip .2in
\noindent
{\bf Acknowledgments}
\vskip .1in
We thank 
Bill Bardeen and Cosmas Zachos for 
helpful discussions. 
This work is supported in part by
the US Department of Energy, High Energy Physics Division,
Contract W-31-109-ENG-38, and 
grant DE-AC02-76CHO3000.

\end{document}